\documentclass{ws-procs975x65_pdflatex}
\usepackage[makeroom]{cancel}

\usepackage{color}

\begin{document}

\title{IMPACT OF NEUTRON DECAY EXPERIMENTS ON NON-STANDARD MODEL PHYSICS}

\author{G. KONRAD$^*$ AND W. HEIL}

\address{Institut f\"ur Physik, Universit\"at Mainz, Staudingerweg 7\\
55099 Mainz, Germany\\
$^*$E-mail: konradg@uni-mainz.de\\
www.quantum.physik.uni-mainz.de}

\author{S. BAE\ss LER$^*$ AND D. PO\v CANI\' C}

\address{Department of Physics, University of Virginia\\
Charlottesville, VA 22904, USA\\
$^*$E-mail: baessler@virginia.edu}

\author{F. GL\"UCK}

\address{IEKP, KIT, 76131 Karlsruhe, Germany\\
KFKI, RMKI, H-1525 Budapest 114, Hungary\\
E-mail: ferenc.glueck@kit.edu}

\begin{abstract}
This paper gives a brief overview of the present and expected future
limits on physics beyond the Standard Model (SM) from neutron beta
decay, which is described by two parameters only within the SM.  Since
more than two observables are accessible, the problem is
over-determined.  Thus, precise measurements of correlations in
neutron decay can be used to study the SM as well to search for
evidence of possible extensions to it.  Of particular interest in this
context are the search for right-handed currents or for scalar and
tensor interactions.  Precision measurements of neutron decay
observables address important open questions of particle physics and
cosmology, and are generally complementary to direct searches for new
physics beyond the SM in high-energy physics.  Free neutron decay is
therefore a very active field, with a number of new measurements
underway worldwide.  We present the impact of recent developments.
\end{abstract}

\keywords{Standard Model; Scalar and tensor interactions; Right-handed
  currents; Neutron beta decay; Neutrino mass; Neutrinoless double
  beta decay} 

\bodymatter

\def\etal{\textit{et al.} }

\section{Introduction}\label{aba:sec1}

Neutron decay, $n\rightarrow p e\overline{\nu}_e$, is the
simplest nuclear beta decay, well described as a purely left-handed,
$V$$-$$A$ interaction within the framework of the Standard Model of
elementary particles and fields.  Thanks to its highly precise
theoretical description\cite{czarnecki:2004}, neutron beta decay
data can lead to limits on certain extensions to the SM.

Neutron decay experiments provide one of the most sensitive means for
determining the weak vector ($L_V G_F V_{ud}$) and axial-vector ($L_A
G_F V_{ud}$) coupling constants, and the element $V_{ud}$ of the
Cabibbo--Kobayashi--Maskawa (CKM) quark-mixing matrix.  Extracted 
$V_{ud}$, along with $V_{us}$ and $V_{ub}$ from K-meson and B-meson 
decays, respectively, test the unitarity of the CKM matrix. $G_F$ is
the Fermi weak coupling constant, evaluated from the muon 
lifetime\cite{amsler:2008}.  The value of $L_V$ is important for
testing the conserved vector current (CVC) hypothesis.  The size of 
the weak coupling constants is important for applications in cosmology (e.g., primordial
nucleosynthesis), astronomy (e.g., solar cycle, neutron star
formation), and particle physics (e.g., neutrino detectors, neutrino
scattering)\cite{abele:2008, nico:2005, dubbers:1991}.

In the SM, the CVC hypothesis requires $L_V=1$ for zero momentum transfer.
Therefore, neutron beta decay is described by two parameters only, $\lambda=L_A/L_V$ and
$V_{ud}$.  The neutron lifetime $\tau_n$ is inversely proportional to
$G_F^2 |V_{ud}|^2(1+3|\lambda|^2)$.  Hence, independent measurements
of $\tau_n$ and of an observable sensitive to $\lambda$, allow the
determination of $V_{ud}$.  The value of $\lambda$ can be determined
from several independent neutron decay observables, introduced in
Sec.~\ref{aba:sec2}.  Each observable brings a different sensitivity
to non-SM physics, such that comparing the various values of $\lambda$
provides an important test of the validity of the SM.  Of particular
interest is the search for scalar and tensor interactions, discussed
in Secs.~\ref{aba:ssec41} and~\ref{aba:ssec42}.  These interactions
can be caused, e.g., by leptoquarks or charged Higgs
bosons.\cite{herczeg:2001} 
In Sec.~\ref{aba:ssec43} we discuss a particular kind of 
$V$$+$$A$ interactions, the manifest left-right symmetric (MLRS) models, 
with the $SU(2)_L \times SU(2)_R \times U(1)_{B-L}$ gauge group
and right-handed charged current, approximately realized with a
minimal Higgs sector.

\section{Measurable parameters of neutron decay}\label{aba:sec2}

The matrix element $\mathcal{M}$ describing neutron beta decay can be
constructed as a four-fermion interaction composed of hadronic and
leptonic matrix elements.  Assuming that vector ($V$), axial-vector
($A$), scalar ($S$), and tensor ($T$) currents are involved, the decay
matrix element can be written as a sum of left-handed and right-handed
matrix elements:
\begin{equation}
  \mathcal{M} = \frac{2 G_F V_{ud}}{\sqrt{2}} \sum_{j \in \{V,A,S,T\}}
  L_j \langle p|\Gamma_j|n\rangle \langle e^-|\Gamma_j
  \frac{1-\gamma_5}{2}|\nu_e\rangle + R_j \langle p|\Gamma_j|n\rangle
  \langle e^-|\Gamma_j \frac{1+\gamma_5}{2}|\nu_e\rangle. 
\label{aba:eq1}
\end{equation}
The four types of currents are defined by the operators:
\begin{equation}
  \Gamma_V = \gamma_\mu , 
  \quad \Gamma_A = i \gamma_\mu\gamma_5 , 
  \quad \Gamma_S = 1 , \quad \text{and} 
  \quad \Gamma_T = \frac{i [\gamma_\mu, \gamma_\nu]}{2\sqrt{2}}.
\label{aba:eq2}
\end{equation}
The coupling constants to left-handed (LH) and right-handed (RH)
neutrinos are denoted by $L_j$ and $R_j$, respectively.  This
parametrization was introduced in Ref.~\refcite{glueck:1995} in order
to highlight the handedness of the neutrino in the participating
$V,A,S,T$ currents. The $L_j$ and $R_j$ coupling constants are linear
combinations of the coupling constants, $C_j$ and $C_j'$, that were 
defined in earlier work\cite{lee:1956}:
\begin{equation}
  C_j = \frac{G_F V_{ud}}{\sqrt{2}}(L_j+R_j), 
  \quad C_j' = \frac{G_F V_{ud}}{\sqrt{2}}(L_j-R_j), 
  \quad \text{for} \quad j = V, A, S, T.
\label{aba:eq3}
\end{equation}

We neglect effects of time-reversal violation, i.e., we consider the
above 8 couplings to be real. 

In neutron decay experiments the outgoing spins are usually not
observed. Summing over these spin quantities, and neglecting the
neutrino masses, one can evaluate the triple differential decay rate
to be\cite{jackson:1957b}: 
\begin{eqnarray}
  {\rm d}^3\Gamma & = & \frac{1}{(2 \pi)^5}\frac{G_F^2 |V_{ud}|^2}{2}
  p_e E_e \left( E_0-E_e \right)^2 {\rm d}E_e {\rm d}\Omega_e {\rm
    d}\Omega_\nu \nonumber \\ 
                  &   & \times \xi \left[ 1+a \frac{\mathbf{p}_e \cdot
      \mathbf{p}_\nu}{E_e E_\nu}+b \frac{m_e}{E_e}+ \mathbf{s}_n
    \left( A \frac{\mathbf{p}_e}{E_e}+B \frac{\mathbf{p}_\nu}{E_\nu} +
    \ldots \right) \right], 
\label{aba:eq4}
\end{eqnarray}
where $p_e$, $p_\nu$, $E_e$, and $E_\nu$ are the electron (neutrino)
momenta and total energies, respectively, $E_0$ is the maximum
electron total energy, $m_e$ the electron mass, $\mathbf{s}_n$ the
neutron spin, and the $\Omega_i$ denote solid angles.  Quantities $a$,
$A$, and $B$ are the angular correlation coefficients, while $b$ is
the Fierz interference term.  The latter, and the neutrino-electron
correlation $a$, are measurable in decays of unpolarized neutrons,
while the $A$ and $B$, the beta and neutrino asymmetry parameters,
respectively, require polarized neutrons.  The dependence of $a$, $b$,
$A$, and $B$ on the coupling constants $L_j$ and $R_j$ is described in
Ref.~\refcite{glueck:1995}.  We mention that in the
presence of LH $S$ and $T$ couplings $B$ depends on the electron
energy: $B=B_0+b_\nu \frac{m_e}{E_e}$, where $b_\nu$ is another
Fierz-like parameter, similar to $b$.\cite{jackson:1957b,glueck:1995}  
We note that $a$, $A$, and $B_0$ are sensitive to non-SM couplings only 
in second order, while $b$ and $b_\nu$ depend in first order on $L_S$ 
and $L_T$.  A non-zero $b$ would indicate the existence of LH $S$ and 
$T$ interactions.

Another observable is $C$, the proton asymmetry relative to the
neutron spin.  Observables related to the proton do not appear in
Eq.~(\ref{aba:eq4}).  However, the proton is kinematically coupled to
the other decay products.  The connection between $C$ and the coupling
constants $L_j$ and $R_j$ is given in Refs.~\refcite{glueck:1995} and
\refcite{glueck:1996}.

We also use the ratio of the $\mathcal{F} t^{0^+ \rightarrow 0^+}$
values in superallowed Fermi (SAF) decays to the equivalent quantity
in neutron decay, $\mathcal{F}t^n$:
\begin{equation}
  r_{\mathcal{F} t} = \frac{\mathcal{F} t^{0^+ \rightarrow
      0^+}}{\mathcal{F} t^n} =\frac{\mathcal{F} t^{0^+ \rightarrow
      0^+}}{f^nt(1+\delta'_R)} =\frac{\mathcal{F} t^{0^+ \rightarrow
      0^+}}{f_R \ln{(2)} \tau_n}, 
\label{aba:eq5}
\end{equation}
where $f^n=1.6887$ is a statistical phase-space
factor\cite{czarnecki:2004}.  The nucleus-dependent (outer) radiative
correction $\delta'_R$, and $\mathcal{O}(\alpha^2)$
corrections\cite{wilkinson:1982,marciano:2006,towner:2008}, change
$f^n$ by $\sim1.5$\,\% to $f_R=1.71385(34)$\footnote{The most recently
published value of $f_R=1.71335(15)$\cite{abele:2004} used
$f^n=1.6886$, and did not include the corrections by Marciano and
Sirlin\cite{marciano:2006}.  Applying the Towner and Hardy
prescription for splitting the radiative corrections\cite{towner:2008}
increases the uncertainty in $f_R$ slightly, to reproduce Eq.~(18) in Ref.~\refcite{marciano:2006}.}.  The corrections implicitly assume
the validity of the $V$$-$$A$ theory.\cite{sirlin:1967} The dependence
of $r_{\mathcal{F}t}$ on coupling constants $L_j$ and $R_j$ is given
in Ref.~\refcite{glueck:1995}.

An electrically charged gauge boson outside the SM is generically
denoted $W'$. The most attractive candidate for $W'$ is the $W_R$
gauge boson associated with the left-right symmetric
models\cite{pati:1974,mohapatra:1975}, which seek to provide a
spontaneous origin for parity violation in weak interactions. $W_L$
and $W_R$ may mix due to spontaneous symmetry breaking.  The physical
mass eigenstates are denoted as
\begin{equation}
  W_1 = W_L \cos \zeta + W_R \sin \zeta, \quad \text{and} \quad
  W_2 = - W_L \sin \zeta +  W_R \cos \zeta,
\label{aba:eq6}
\end{equation}
where $W_1$ is the familiar $W$ boson, and $\zeta$ is the mixing angle
between the two mass eigenstates.  In the MLRS model, there are only
three free parameters, the mass ratio $\delta=m_1^2/m_2^2$, $\zeta$,
and $\lambda'$, while $m_{1,2}$ denote the masses of $W_{1,2}$,
respectively.  Since $L_V=1$ (CVC) and $L_S=L_T=R_S=R_T=0$, the
coupling constants $L_A$, $R_V$, and $R_A$ depend on $\delta$,
$\zeta$, and $\lambda'$ as described in
Refs.~\refcite{beg:1977,glueck:1995}.  The dependence of $a$, $A$,
$B$, $C$, and $\tau_n$ on $\delta$, $\zeta$, and $\lambda'$ follows
from their respective dependence on $L_V$, $L_A$, $R_V$, and $R_A$.

\section{Experimental Data}\label{aba:sec3}

We present results of least-squares fits, using recent experimental
data as well as target uncertainties for planned experiments on
neutron decay.  The principle of non-linear $\chi^2$ minimization is
discussed, e.g., in Ref.~\refcite{eadie:1971}.
Figures~\ref{aba:fig1}--\ref{aba:fig6} show the present and expected
future limits from neutron decay, respectively.  The confidence regions
in 2 dimensions, or confidence intervals in 1 dimension, are defined as
in Ref.~\refcite{press:2007}.

We first analyze the presently available data on neutron decay.  As
input for our study we used: $a=-0.103(4)$ and $B=0.9807(30)$ (both 
from Ref.~\refcite{amsler:2008}), as well as
$\mathcal{F}t^{0^+ \rightarrow 0^+}=3071.81(83)$\,s as the average
value for SAF decays (from Ref.~\refcite{hardy:2009}).  We used our
own averages for $\tau_n$ and $A$, as follows.

The most recent result of Serebrov {\etal\cite{serebrov:2005b}},
$\tau_n=878.5(8)$\,s, is not included in the PDG 2008 average.  We
prefer not to exclude this measurement without being convinced that it
is wrong, and include it in our average to obtain
$\tau_n=(881.8\pm1.4)$\,s.  Our average includes a scale factor of
2.5, as we obtain $\chi^2=45$ for 7 degrees of freedom.  The
statistical probability for such a high $\chi^2$ is
$1.5\times10^{-7}$.  If our average were the true value of the neutron
lifetime $\tau_n$, both the result of Serebrov \etal and the PDG
average would be wrong at the $2-3$\,$\sigma$ level.

Two beta asymmetry experiments have completed their analyses since the
PDG 2008 review.  The UCNA collaboration has published
$A=-0.1138(46)(21)$.\cite{pattie:2009} The last PERKEO II run has
yielded a preliminary value of $A=-0.1198(5)$.\cite{abele:2010}.  We
include these new results in our average, and obtain $A=-0.1186(9)$,
which includes a scale factor of 2.3 based on $\chi^2=28$ for 5
degrees of freedom.  The statistical probability for such a high
$\chi^2$ is $5\times10^{-5}$, not much better than in the case of
$\tau_n$.

Hence, we find that the relative errors are about 4\,\% in $a$, 1\,\%
in $A$, and 0.3\,\% in $B$.  We will not use 
$C=-0.2377(26)$\cite{amsler:2008} in the analysis of present results, 
since the PERKEO II results for $B$ and $C$ are derived from the same 
data set.\footnote{We note that a recent experiment\cite{kozela:2009} 
measured the neutron spin--electron spin correlation $N$ in neutron 
decay. $N$ is the coefficient of an additional term $\left(+N
\mathbf{s}_n \mathbf{s}_e\right)$, which appears in Eq.~(\ref{aba:eq4}) 
if the electron spin is detected.  The $N$ parameter depends linearly 
on $S,T$ couplings. We disregard the result, as it lacks the precision 
to have an impact on our analysis.}

About a dozen new instruments are currently planned or under
construction.  For recent reviews {see
  Refs.~\refcite{abele:2008,nico:2009}}.  We will discuss a future
scenario which assumes the following improvements in precision in a couple of years.
\begin{itemlist}
	\item 
	$\Delta a/a = 0.1$\,\%: Measurements of the neutrino-electron correlation coefficient $a$ with the
          {\textit{a}SPECT\cite{glueck:2005, baessler:2008}},
          {aCORN\cite{wietfeldt:2009}}, {Nab\cite{pocanic:2009}}, and
          PERC experiments are projected or underway. 
	\item
	$\Delta b = 3 \times 10^{-3}$: The first ever measurement of
          the Fierz interference term $b$ in neutron decay is planned
          by the Nab collaboration\cite{pocanic:2009}.  In
          addition, the UCNb\cite{hickerson:2009} and PERC
          collaborations are exploring measurements of $b$. 
	\item
	$\Delta A/A = 3 \times 10^{-4}$:
          Measurements of the beta asymmetry parameter $A$ with PERKEO
          {III\cite{maerkisch:2009}}, {UCNA\cite{plaster:2008}},
          {abBA\cite{alarcon:2007a}}, and PERC\cite{dubbers:2008}
          are either planned or underway.
	\item
	$\Delta B/B = 0.1$\,\%: The abBA\cite{alarcon:2007a} and
          UCNB\cite{wilburn:2009} collaborations intend to measure the
          neutrino asymmetry parameter $B$.  PERC is also
          exploring a measurement of $B$.
	\item
	$\Delta C/C = 0.1$\,\%: 
          The \textit{a}SPECT\cite{zimmer:2010} and PANDA\cite{PANDA} 
          collaborations plan measurements of the proton asymmetry parameter
          $C$; PERC may follow suit as well. 
	\item
	$\Delta \tau_n = 0.8$\,s: Measurements of the neutron lifetime
          $\tau_n$ with beam
          {experiments\cite{dewey:2009,shimizu:2009}}, {material
            bottles\cite{arzumanov:2009,serebrov:2009b}}, and magnetic
          storage
          {experiments\cite{walstron:2009,ezhov:2009b,materne:2009,leung:2009,shaughnessy:2009}} 
          are planned or underway. 
\end{itemlist}
Our assumptions about future uncertainties for $a$, $A$, $B$, and $C$
reflect the goal accuracies in the proposals, while for $\tau_n$ we
only assume the present discrepancy to be resolved.  Our assumed
$\Delta\tau_n$ corresponds to the best uncertainty claimed in a
previous experiment\cite{serebrov:2005b}.

Our scenario ``future limits'' assumes that the SM holds and connects
the different observables.  We used $a=-0.10588$, $b=0$, $B=0.98728$,
$C=-0.23875$, and $\tau_n=882.2$\,s derived from $A=-0.1186$ and
$\mathcal{F} t^{0^+ \rightarrow 0^+}=3071.81$\,s.  These values agree
with the present measurements within 2\,$\sigma$.

\section{Searches for physics beyond the Standard Model}\label{aba:sec4}

Our fits are not conclusive if all 8 coupling constants $L_j$ and
$R_j$, for $j=V,A,S,T$, are treated as free parameters.  We are more
interested in restricted analyses presented below.  Experiments quote
$a$, $A$, $B$, and $C$ after applying (small) theoretical corrections
for recoil and radiative effects; we neglect any dependence on non-SM
physics in these corrections.

\subsection{Left-handed S, T currents}\label{aba:ssec41}

Addition of LH $S$, $T$ currents to the SM leaves $L_V = 1$, $L_A =
\lambda$, $L_S$, and $L_T$ as the non-vanishing parameters.  Non-zero
Fierz interference terms $b$ and $b_\nu$ appear in this model; the
direct determination of $b$ through beta spectrum shape measurement is
the most sensitive way to constrain the size of the non-SM currents.
The experiments discussed above measure the correlation coefficients
from the electron spectra and asymmetries, respectively.  The
published results on $a$, $A$, $B$, and $C$ assume $b = b_\nu = 0$.
To make use of measured values of $a$ in a scenario involving a
non-zero value for the Fierz term $b$, we rewrite Eq.~(\ref{aba:eq4})
for unpolarized neutron decay:
\begin{equation}
  {\rm d}^3\Gamma \propto \left(1+a \frac{\mathbf{p}_e \cdot
  \mathbf{p}_\nu}{E_e E_\nu}+b \frac{m_e}{E_e}\right) 
  \approx \left(1+b
  m_e \left<E_e^{-1}\right>\right)\left(1+\frac{a}{1+b m_e
  \left<E_e^{-1}\right>} \frac{\mathbf{p}_e \cdot \mathbf{p}_\nu}{E_e
  E_\nu}\right). 
\label{aba:eq7}
\end{equation}
The value quoted for $a$ is then taken as a measurement of $\bar
a$, defined through 
\begin{equation}
	\bar{a} = \frac{a}{1+b m_e \left<E_e^{-1}\right>},
\label{aba:eq8}
\end{equation}
where $\left<\cdot\right>$ denotes the weighted average over the part
of the beta spectrum observed in the particular experiment.  This
procedure has been also applied in
Refs.~\refcite{glueck:1995,severijns:2006,hardy:2009}.  Reported
experimental values of $A$, $B$, and $C$ are interpreted as
measurements of
\begin{equation}
  \bar{A} = \frac{A}{1+b m_e \left<E_e^{-1}\right>}, \quad 
  \bar{B} = \frac{B_0+b_\nu m_e \left<E_e^{-1}\right>}{1+b m_e
  \left<E_e^{-1}\right>}, \quad  
  \bar{C} = \frac{-x_C(A+B_0)-x_C' b_\nu}{1+b m_e\left<E_e^{-1}\right>}, 
\label{aba:eq9}
\end{equation}
where $x_C=0.27484$ and $x_C'=0.1978$ are kinematical
factors, assuming integration over all electrons.\footnote{Note that we define $C$ with the opposite sign
compared to Ref. \refcite{glueck:1995} to adhere to the convention
that a positive asymmetry indicates that more particles are emitted in
the direction of spin.}  This procedure is not perfect.  The presence
of a Fierz term $b$ might influence systematic uncertainties.  For
example, the background estimate in PERKEO II assumes the SM
dependence of the measured count rate asymmetry on $E_e$.  The term $m_e
\left<E_e^{-1}\right>$ depends on the part of the electron spectrum
used in each experiment.  We have used the following values in
our study: $m_e \left<E_e^{-1}\right>=0.5393$ for $\bar{A}$, dominated
by PERKEO {II\cite{abele:2002}}, $m_e \left<E_e^{-1}\right>=0.6108$ for $\bar{B}$, dominated by Serebrov
\etal\cite{serebrov:1998a,serebrov:1998b} and PERKEO
{II\cite{schumann:2007c}}, and the mean value $m_e
\left<E_e^{-1}\right>=0.6556$, taken over the whole beta spectrum, for $\bar{a}$ and $\bar{C}$.

\begin{figure}[htb]%
\begin{minipage}[htb]{5.7cm}
	\centering
	\includegraphics[width=5.7cm]{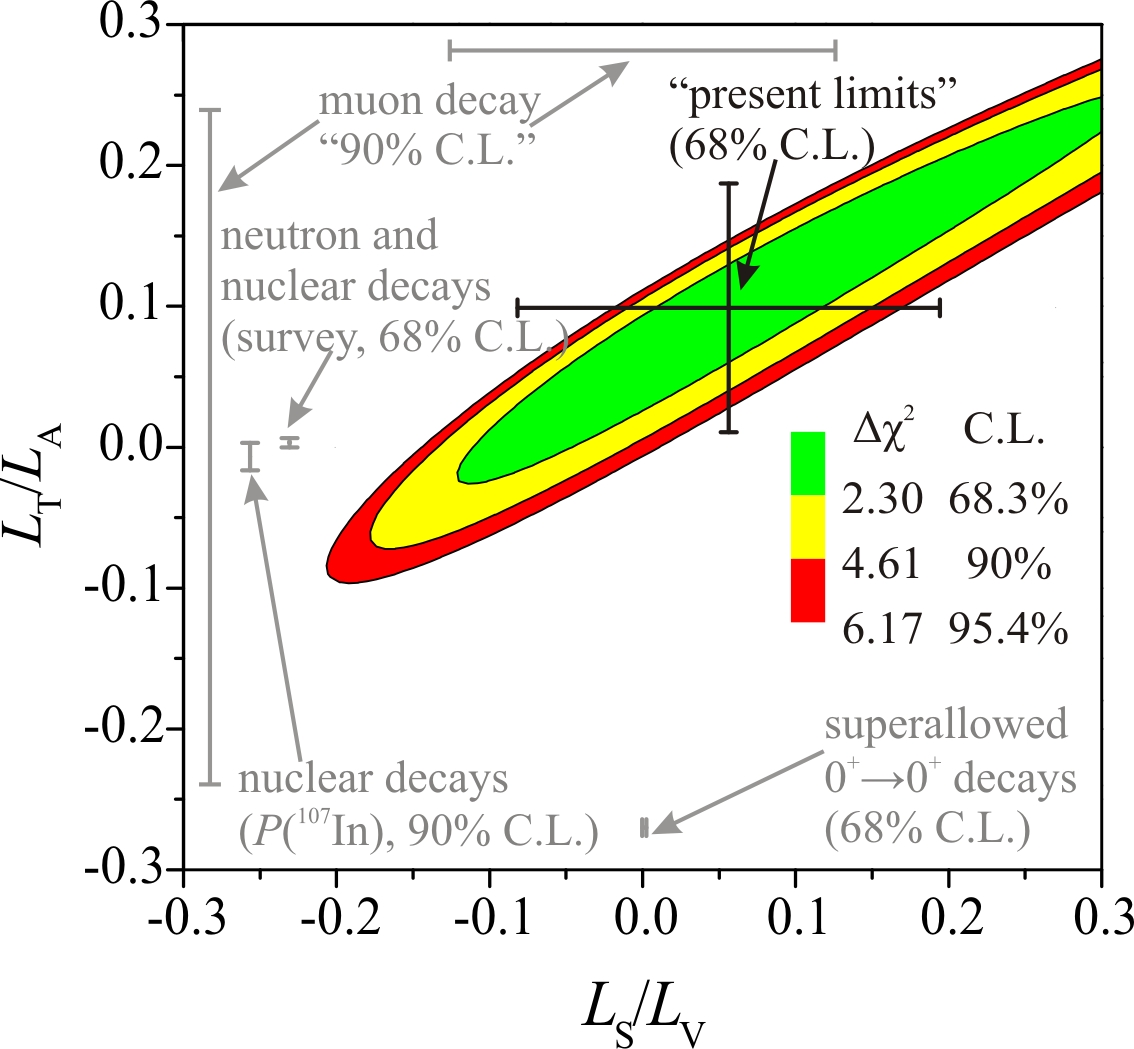}
	\caption{Present limits from neutron decay (only $a$, $A$, and
	$B$).  The SM values are at the origin of the plot.  Analogous
	limits extracted from muon decays are indicated.  Other limits
	are discussed in the text.  All bars correspond to single
	parameter limits.}
	\label{aba:fig1}
\end{minipage}
\hfill
\begin{minipage}[htb]{5.7cm}
	\centering
	\includegraphics[width=5.7cm]{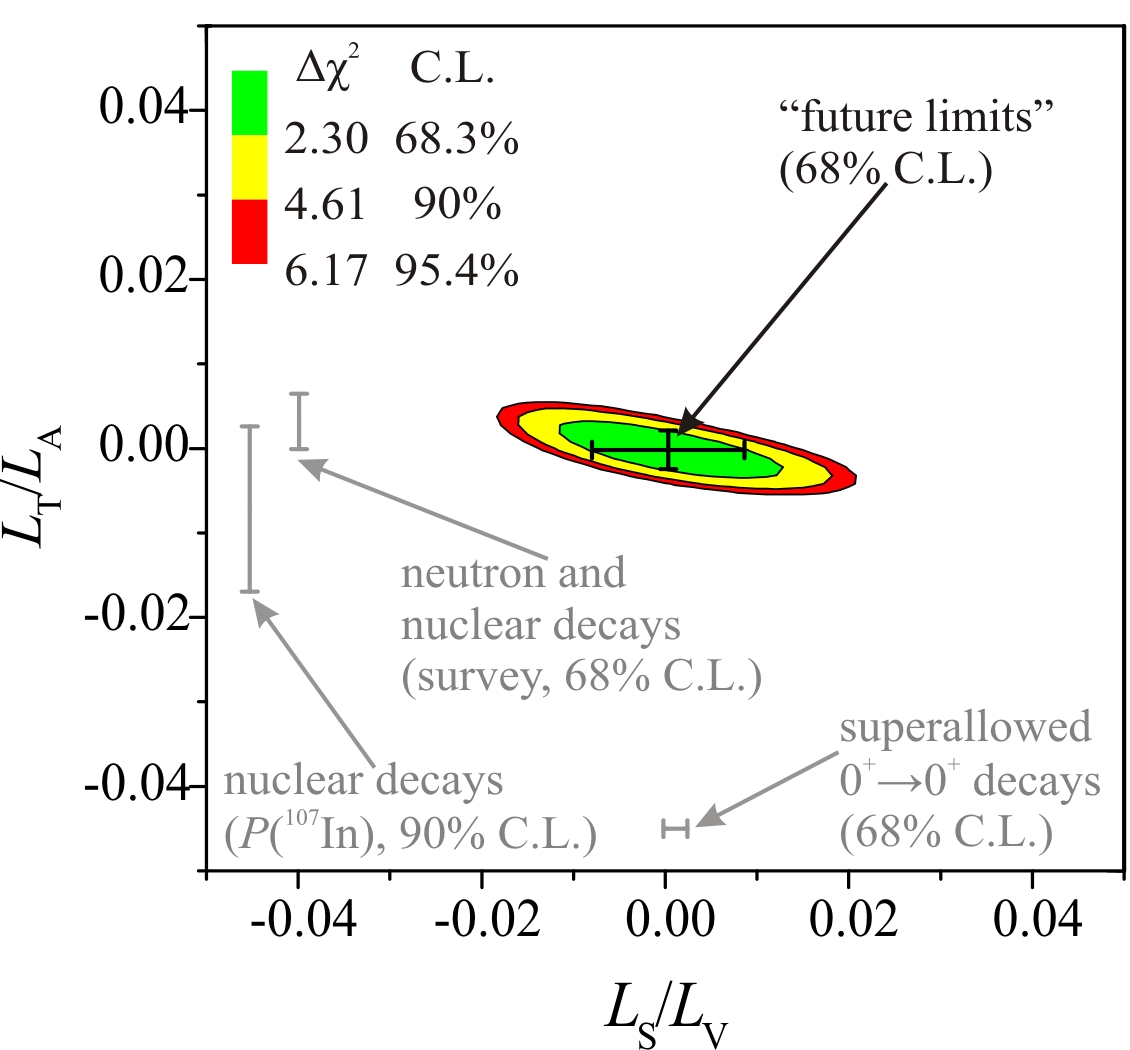}
	\caption{Future limits from neutron decay, assuming improved and independent
	measurements of $a$, $b$, $A$, $B$, and $C$.  Analogous limits
	extracted from muon decays are not indicated since they exceed
	the scale of the plot.}
	\label{aba:fig2}
\end{minipage}
\end{figure}

Figure~\ref{aba:fig1} shows the current limits from neutron decay.
Free parameters $\lambda$, $L_S/L_V$, and $L_T/L_A$ were fitted to the
observables $a$, $b$, $A$, $B$, and $C$.  Unlike
Secs.~\ref{aba:ssec42} and~\ref{aba:ssec43}, here we omit the neutron
lifetime $\tau_n$, since otherwise we would have to determine the
possible influence of the Fierz term in Fermi decays, $b_{\rm F}$, on
the $\mathcal{F}t^{0^+ \rightarrow 0^+}$ values.  A combined analysis
of neutron and SAF beta decays will be published soon.

Figure~\ref{aba:fig2} presents the impact of projected measurements in
our future scenario.  For comparison, a recent combined analysis of
nuclear and neutron physics data (see Ref.~\refcite{severijns:2006})
finds $L_S/L_V = 0.0013(13)$ and $L_T/L_A = 0.0036(33)$, with
1\,$\sigma$ statistical errors.  It includes the determination of the
Fierz term $b_F$ from superallowed beta decays, updated in
Ref.~\refcite{hardy:2009}, which sets a limit on $L_S$ that is hard to
improve with neutron decay alone.  As in the recent survey of
Severijns \etal\cite{severijns:2006}, we do not include the limits on
tensor couplings obtained\cite{boothroyd:1984} from a measurement of
the Fierz term $b_{GT}$ in the forbidden Gamow-Teller decay of
$^{22}$Na, due to its large $\log ft($=7.5$)$ value.  Neutron decay
has the potential to improve the best remaining nuclear limit on $L_T$
as provided by a measurement of the longitudinal polarization of
positrons emitted by polarized $^{107}$In nuclei ($\log
ft=5.6$)\cite{camps:1997,severijns:2000}.  Limits from neutron decay
are independent of nuclear structure.  The stringent limit on $L_T$ in
Ref.~\refcite{severijns:2006} stems mainly from measurements of
$\tau_n$ and $B$ in neutron decay.  New neutron decay experiments
alone could lead to an accuracy of $\Delta(L_T/L_A)=0.0023$,
competitive with the combined analysis of neutron and nuclear physics
data\cite{severijns:2006}, and $\Delta(L_S/L_V)=0.0083$, both at the
1\,$\sigma$ confidence level.  Supersymmetric (SUSY) contributions to
the SM can be discovered at this level of precision, as discussed in
Ref.~\refcite{profumo:2007}.

\subsection{Right-handed S, T currents}\label{aba:ssec42}

\begin{figure}[htb]%
\begin{minipage}[htb]{5.7cm}
	\centering
	\includegraphics[width=5.7cm]{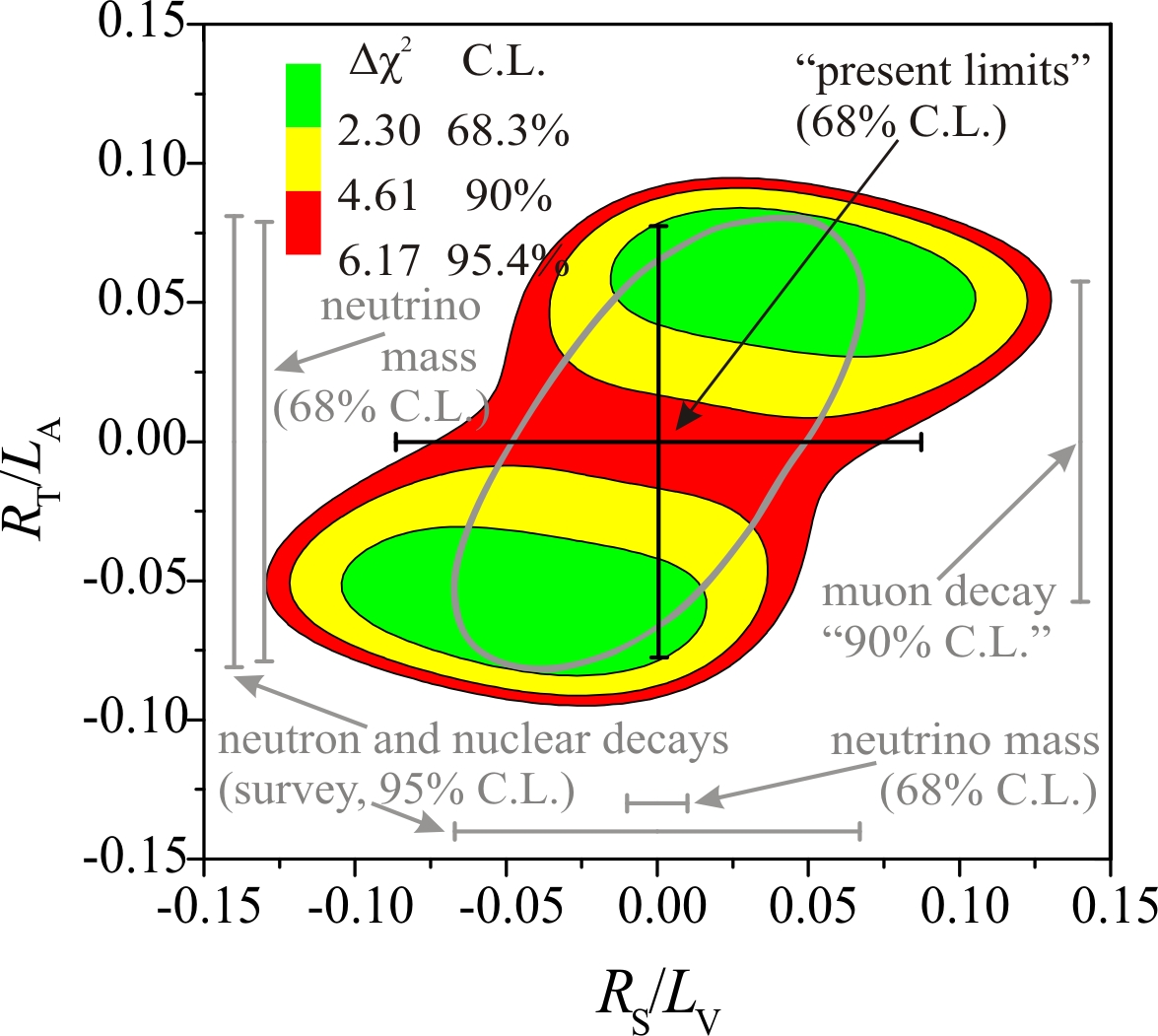}
	\caption{Current limits from $a$, $A$, $B$, and $\tau_n$ in
	neutron decay.  The SM prediction is at plot
	origin. As a comparison, we show limits from a survey of nuclear and
	neutron beta decays\cite{severijns:2006}, and limits from muon 
	decays and neutrino mass measurements.  The grey
	ellipse is the present 86.5\,\% contour from Ref.~\refcite{severijns:2006}.  The muon limit on $R_S/L_V$ 
	is larger than the scale of the plot.}
	\label{aba:fig3}
\end{minipage}
\hfill
\begin{minipage}[htb]{5.7cm}
	\centering
	\includegraphics[width=5.7cm]{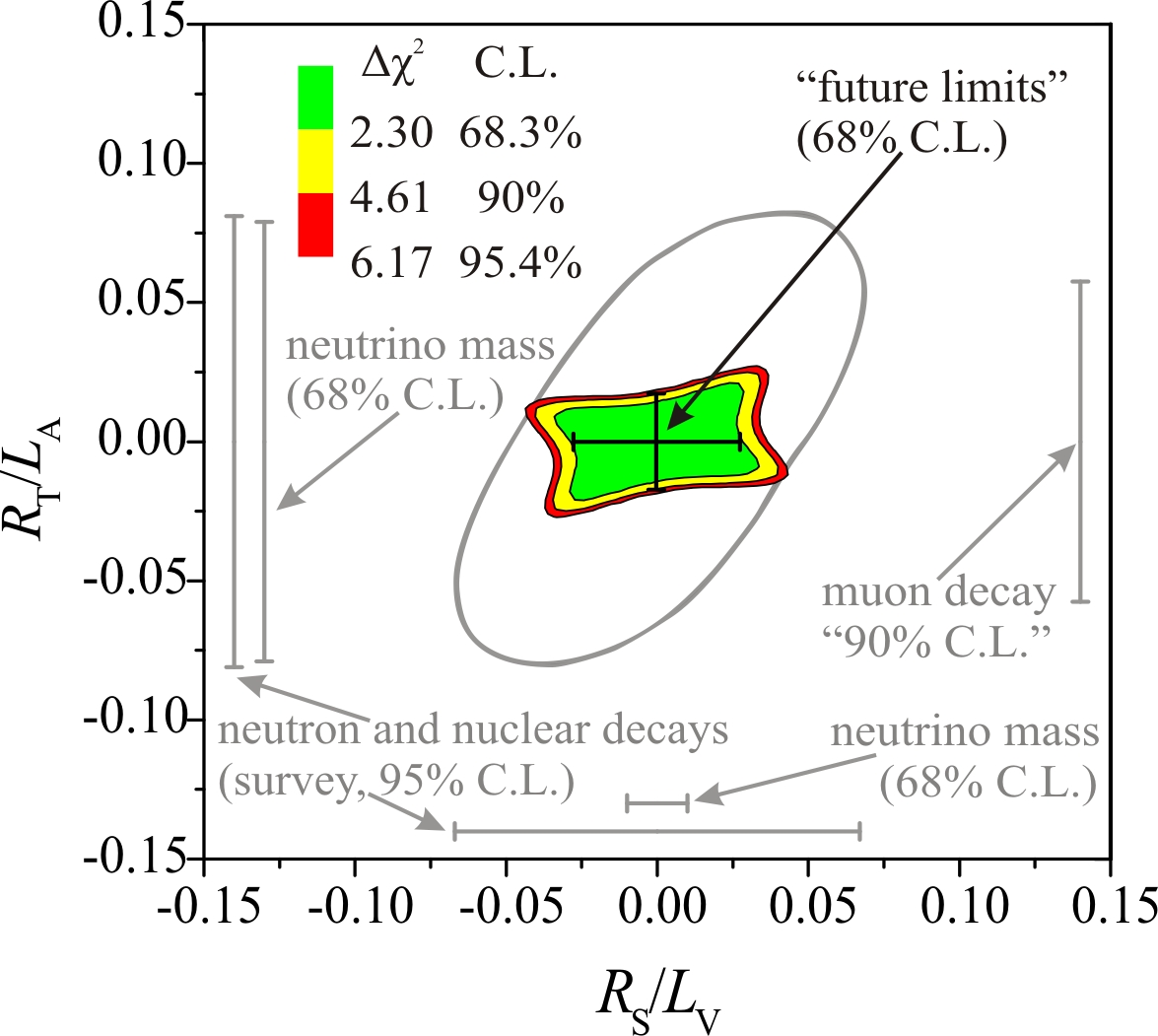}
	\caption{Future limits from neutron decay, assuming improved
	measurements of $a$, $A$, $B$, $C$, and $\tau_n$.   The grey
	ellipse is the present 86.5\,\% contour from a recent survey of nuclear and
	neutron beta decays\cite{severijns:2006}.
	Analogous tensor limits from muon decays are also shown---the
	scalar limits are larger than the scale of the plot (details
	in text).  \newline}
	\label{aba:fig4}
\end{minipage}
\end{figure}

Adding the RH $S$ and $T$ currents to the SM yields $L_V = 1$, $L_A =
\lambda$, $R_S$, and $R_T$ as the remaining non-zero parameters.  The
observables depend only quadratically on $R_S$ and $R_T$, i.e., the
possible limits are less sensitive than those obtained for LH $S$, $T$
currents.  Figure~\ref{aba:fig3} shows the present limits from neutron
decay.  A similar analysis of this scenario was recently published in
Ref.~\refcite{schumann:2007b}.

Free parameters $\lambda$, $R_S/L_V$, and $R_T/L_A$ were fitted to the
observables $a$, $A$, $B$, $C$, and $\tau_n$.  Additionally, to take
into account uncertainties in the $\mathcal{F} t$ values and in
radiative corrections, we fitted $\mathcal{F}t^{0^+ \rightarrow 0^+}$
and $f^R$ to `data points' $3071.81(83)$\,s and $1.71385(34)$, respectively.

The Fierz interference terms $b$ and $b_\nu$ are zero in this model.
Hence, measurements of $b$ (or $b_F$ in SAF beta decays) can invalidate the model, but not determine
its parameters.  Figure~\ref{aba:fig4} shows the
projected improvement in our future scenario.  The grey ellipse stems
from a recent survey of the state of the art in nuclear and neutron
beta decays\cite{severijns:2006}.  New neutron decay experiments
alone could considerably improve the limits on RH $S$ and $T$
currents, to $\Delta(R_S/L_V)=0.0275$ and $\Delta(R_T/L_A)=0.0173$.

\subsection{Right-handed W bosons}\label{aba:ssec43}

\begin{figure}[htb]%
\begin{minipage}[htb]{5.7cm}
	\centering
	\includegraphics[width=5.7cm]{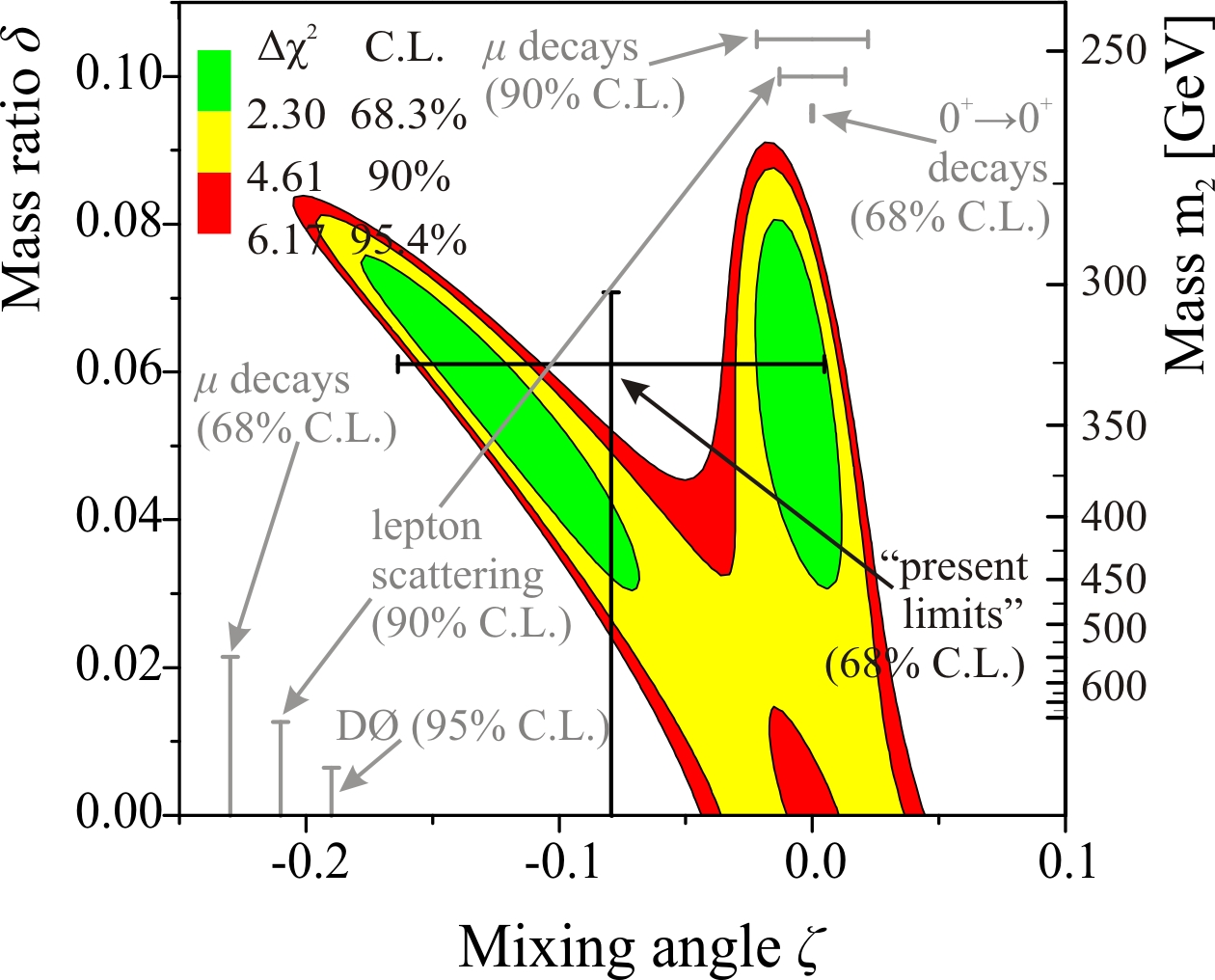}
	\caption{Current limits from $a$, $A$, $B$, and $\tau_n$ in
	neutron decay.  The SM prediction is at plot origin. As a 
	comparison, we show analogous limits from muon {decays
	\cite{barenboim:1997,macdonald:2008}}, lepton scattering (deep 
	inelastic $\nu$-hadron, $\nu$-$e$ scattering and $e$-hadron 
	interactions{)\cite{czakon:1999}}, and a direct search at {D$\cancel{0}$\cite{abazov:2008}}.}
	\label{aba:fig5}
\end{minipage}
\hfill
\begin{minipage}[htb]{5.7cm}
	\centering
	\includegraphics[width=5.7cm]{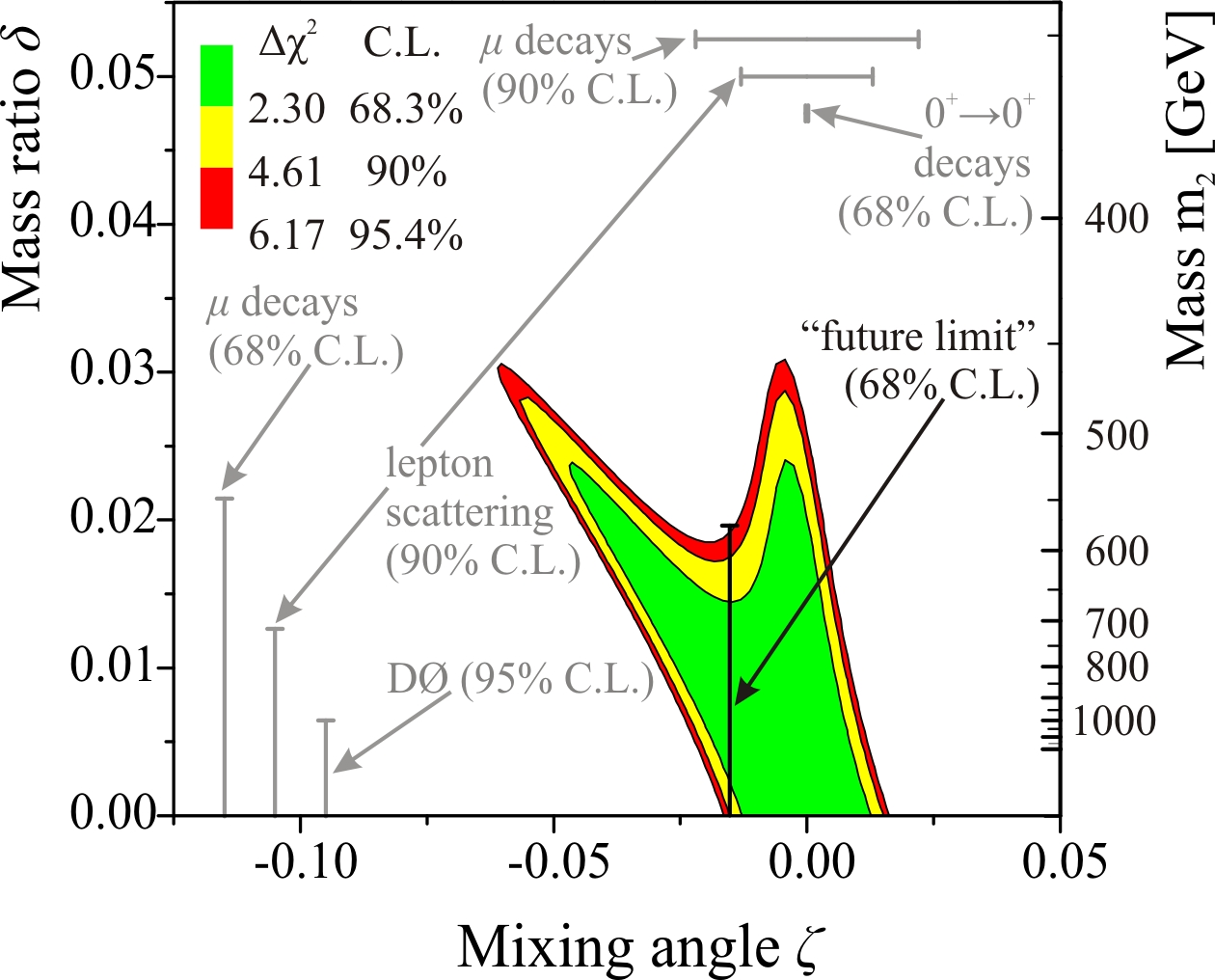}
	\caption{Projected future limits from neutron decay, assuming
	improved measurements of $a$, $A$, $B$, $C$, and $\tau_n$.
	The value of $|V_{ud}|$ from superallowed $0^+\to 0^+$ nuclear
	beta decays was used to set a limit on $\zeta$, assuming that
	the CKM matrix for LH quarks is strictly unitary (see
	Ref.~\refcite{hardy:2009}).}
	\label{aba:fig6}
\end{minipage}
\end{figure}

Adding RH $V$ and $A$ currents to the SM leaves $\delta$, $\zeta$, and
$\lambda^\prime$ as the non-vanishing parameters.
Figure~\ref{aba:fig5} shows the current limits from neutron decay.
The fit parameters $\delta$, $\zeta$, $\lambda^\prime$,
$\mathcal{F}t^{0^+ \rightarrow 0^+}$, and $f^R$, were fitted to the
observables discussed in Sec.~\ref{aba:ssec42}.

Measurements of the polarized observables, i.e., the electron,
neutrino, or proton asymmetries, lead to important restrictions, but
are at present inferior to limits on the mixing angle $\zeta$ from 
$\mu$ decays\cite{macdonald:2008}.  They are also inferior to limits
on the mass $m_2$ from direct searches for extra $W$
bosons\cite{amsler:2008}.  Comparison of beta decay limits with high
energy data is possible in our minimal MLRS model.  For example, the
comparison with $W^\prime$ searches at Tevatron\cite{abazov:2008}
assumes a RH CKM matrix identical to the LH one and identical
couplings.  In more general scenarios the limits are complementary to
each other since they probe different combinations of the RH
parameters\cite{severijns:1998}.

Figure~\ref{aba:fig6} shows the improvement from planned measurements
in our future scenario.  The $\chi^2$ minimization converges to a
single minimum at mass $m_2=\infty$; with $\chi^2=0$, i.e., the mixing
angle $\zeta$ is not defined at this minimum.  The 68.3\,\% C.L. is
$\delta<0.0196$ which yields $m_2>574$\,GeV.  In the mass range
$>1$\,TeV, not excluded by collider experiments, we would improve the
limit on $\zeta$ from $\mu$ decays slightly.

We emphasize that all presented RH coupling limits ($R_S$, $R_T$,
$\delta$, and $\zeta$) assume that the RH (Majorana) neutrinos are
light ($m\ll 1\,$MeV). The RH interactions are kinematically weakened
by the masses of the predominantly RH neutrinos, if these masses are not much
smaller than the electron endpoint energy in neutron decay (782 keV).
If both the $W$ boson and neutrino left-right mixing angles were zero, and if the
RH neutrino masses were above 782 keV, RH corrections to neutron decay
observables would be completely absent.

In summary, new physics may be within reach of precision measurements
in neutron beta decay in the near future.

\section{Limits from other measurements}\label{aba:sec5}

\subsection{Constraints from muon and pion decays}\label{aba:ssec51}

Muon decay provides arguably the theoretically cleanest limits on
non-($V$$-$$A$) weak interaction
couplings\cite{amsler:2008,gagliardi:2005}. Muon decay involves
operators that are different from the ones encountered in neutron, and
generally hadronic, decays. However, in certain models (e.g., the SUSY 
extensions discussed in Ref.~\refcite{profumo:2007}, or in the MLRS), 
the muon and neutron decay derived limits become
comparable\cite{cirigliano:2009}. In order to
illustrate the relative sensitivities of the muon and neutron sectors,
we have attempted to translate the muon limits from
Refs.~\refcite{amsler:2008} and \refcite{gagliardi:2005} into
corresponding neutron observables such as $L_S/L_V$, $L_T/L_A$, and
$R_T/R_A$. In doing so we neglected possible differences in SUSY
contributions to muon and quark decays, making the comparison merely
illustrative. These limits are plotted in
Figs.~\ref{aba:fig1}--\ref{aba:fig6}, as appropriate, showing that
neutron decay measurements at their current and projected future
sensitivity are not only complementary, but also competitive to the
muon sector. 

Limits similar to the ones discussed in Sec.~\ref{aba:sec5} can be
extracted from pion decays (added complexity of heavier meson decays
limits their sensitivity). The presence of a tensor interaction would
manifest itself in both the Fierz interference term in beta decays
(e.g., of the neutron) and in a non-zero value of the tensor form
factor for the pion. The latter was hinted at for well over a decade,
but was recently found to be constrained to $-5.2 \times 10^{-4} < F_T
< 4.0 \times 10^{-4}$ with 90\,\% C.L.\cite{bychkov:2009} While
values for $b$ in neutron decay and for the pion $F_T$ are not
directly comparable, in certain simple scenarios they would be of the
same order\cite{herczeg:2004}. Thus, finding a non-zero value for
$b$ in neutron decay at the level of $\mathcal{O}(10^{-3})$ would be
extremely interesting. Similarly, the $\pi \rightarrow e \nu$ decay
($\pi_{e2}$) offers a very sensitive means to study non-($V$$-$$A$)
weak couplings, primarily through a pseudoscalar term in the
amplitude. Alternatively, $\pi_{e2}$ decay provides the most sensitive
test of lepton universality. Thus, new measurements in neutron decay
would complement the results of precision experiments in the pion
sector, such as PIBETA\cite{PIBETA} and {PEN\cite{PEN}}. 

\subsection{RH coupling constraints from $0\nu$ double
  $\beta$ decay, and $m_\nu$}\label{aba:ssec52}

The most natural mechanism of neutrinoless ($0\nu$) double beta decay
is through virtual electron neutrino exchange between the two neutron
decay vertices.  The LH and RH $\nu_e$ may mix with mass eigenstate
Majorana neutrinos $N_i$\cite{doi:1985}:

\begin{equation}
\nu_{eL}=\sum_{i=1}^6 U_{ei} \frac{1-\gamma_5}{2} N_i, \quad
\text{and} \quad \nu_{eR}=\sum_{i=1}^6 V_{ei} \frac{1+\gamma_5}{2}
N_i,
\end{equation}
where $U_{ei}$ and $V_{ei}$ denote elements of the LH and RH mixing
matrices, respectively.

The neutrinoless double beta decay amplitude with the virtual neutrino
propagator has two parts\cite{paes:1999}.  If the SM LH $V$$-$$A$
coupling combines with LH coupling terms (LL interference), the
amplitude contribution is proportional to the Majorana neutrino masses
(weighted with the $U_{ei}^2$ factors). Since from neutrino oscillations 
we have rather small lower limits for these masses ($40$\,meV for the 
heaviest LH neutrino\cite{mohapatra:2006}), we get  only weak 
constraints for the non-SM LH couplings.
On the other hand, if the SM LH $V$$-$$A$ coupling combines with RH
non-SM terms (LR interference), the amplitude is proportional to the
virtual neutrino momentum (instead of the neutrino mass); since the
momentum can be quite large we get constraints for the RH non-SM
couplings.  The latter part of the $0\nu$ double beta decay amplitude
is proportional to the effective RH couplings $\tilde R_j=R_j
\varepsilon$, for $j=V,A,S,T$, where\cite{doi:1985,hirsch:1996}
\begin{equation}
  \varepsilon=\sum_{i=1}^6{}^{(\text{light})}  U_{ei} V_{ei}, \quad
    \text{where ``light'' implies}\ m_i<10\,\text{MeV}.
\end{equation}

According to Ref.~\refcite{doi:1985} there are three different
scenarios:
\begin{itemlist}[M-II:]
\item[D:] all neutrinos are light Dirac particles $\Longrightarrow$ no
constraints for non-SM couplings because $\varepsilon=0$.

\item[M-I:] all neutrinos are light ($<1$\,MeV) Majorana particles
  $\Longrightarrow$ no constraints for non-SM couplings, because
  $\varepsilon=0$ from orthogonality condition. 

\item[M-II:] both light ($<$\,MeV) and heavy ($>$\,GeV)
  Majorana neutrinos exist $\Longrightarrow$ constraints for non-SM
  couplings: $\varepsilon \neq 0$, because heavy neutrinos are missing
  from the sum; $\varepsilon$ is on the order of the unknown, likely
  small, mixing angle $\theta_{LR}$ between LH and RH neutrinos.

\end{itemlist}

In the M-II scenario there are stringent constraints for the effective
RH $V,A,S,T$ couplings: $|\tilde R_j|<10^{-8}$\cite{paes:1999}. These effective couplings are proportional
to $\varepsilon \sim \theta_{LR}$\cite{doi:1985,hirsch:1996}.  Since $\varepsilon$ depends on
specific neutrino mixing models, it is not possible to give model
independent limits for the $R_j$ couplings based on $0\nu$
double $\beta$ decay data. 
We have already mentioned in Sec.~\ref{aba:ssec43} that for the heavy RH 
(Majorana) neutrinos the RH observables in neutron decay are kinematically 
weakened or for special cases completely suppressed.

Assuming 1\,TeV effective RH neutrino mass scale within M-II, one
obtains $|\zeta|<4.7 \times 10^{-3}$ and $m_2>1.1$\,TeV\cite{hirsch:1996}. 
 For a larger RH neutrino mass scale
these constraints become weaker.

In Ref. \refcite{klapdor:2006} it is argued that neutrinoless double beta 
decay occurs in nature. If further experiments confirm this observation, 
one can be sure that the neutrinos are Majorana particles.

The RH couplings can contribute to neutrino mass through loop effects,
leading to constraints on the RH coupling constants from neutrino mass
limits\cite{ito:2005}.  Using the absolute neutrino mass limit
$m(\nu_e)<2.2$\,eV from the Troitsk and Mainz tritium decay
experiments\cite{lobashev:2003,kraus:2005}, one obtains the
1\,$\sigma$ limits: $|R_S|<0.01$, $|R_T|<0.1$, and $|R_V-R_A|<0.1$.
With the $m(\nu_e)<0.22$\,eV model dependent limit from
cosmology\cite{komatsu:2009}
(similar neutrino mass limit is expected from the KATRIN
experiment\cite{angrik:2004}), the above coupling constant limits
become 10 times more restrictive. An intermediate neutrino mass upper limit of
order $0.5-0.6$\,eV comes from neutrinoless double beta decay\cite{klapdor:2006}
and from other cosmology {analysis\cite{tegmark:2004}}.

\section*{Acknowledgements}\label{aba:sec6}
This work was supported by the German Federal Ministry of Education and Research under Contract No. 06MZ989I, 06MZ170, the European Commission under Contract No. 506065, the Universit\"at Mainz, and the National Science Foundation Grants PHY-0653356, -0855610, and -0970013.

\bibliographystyle{ws-procs975x65}
\bibliography{bibliography}

\end{document}